\documentclass{wileyarticle}

\articletype{}%Article Type}%
\artid{6384759}%
\jid{}%
\jvol{201}
\jyear{2017}%
\jissue{2}%
\jmonth{March}%
\doi{10.1155/2015/6384759}%
\copyeditor{}%
\startpage{1}%
\received{26 April 2016}
\revised{6 June 2016}
\accepted{6 June 2016}

\usepackage{lipsum}% this package is included to get dummy paragraphs for sample purpose.
\usepackage{graphicx,pstricks,epsfig,rotating,amsmath,amssymb}

\newcommand{\bea}{\begin{eqnarray}}
\newcommand{\eea}{\end{eqnarray}}
\newcommand{\apgt} {\ {\raise-.5ex\hbox{$\buildrel>\over\sim$}}\ }

\begin{document}

\title{Mixed Phase within the Multi-polytrope Approach to High Mass Twins.}

\author[1,2]{David Alvarez-Castillo}
\author[1,3,4]{David Blaschke}
\author[5,6]{Stefan Typel}

%\author[2,3]{Author Two\footnotemark{3}}

%\author[3]{Author Three}

\authormark{ALVAREZ-CASTILLO \textsc{et al}}

%\footnotetext[2]{This is an example for first author footnote.}
%\footnotetext[3]{This is an example for second author footnote.}
%\footnotetext[4]{Abbreviations: ANA, anti-nuclear antibodies; APC, antigen-presenting cells; IRF, interferon regulatory factor}

\address[1]{\orgdiv{Bogoliubov  Laboratory of Theoretical Physics}, \orgname{Joint Institute for Nuclear Research}, \orgaddress{\state{Joliot-Curie str. 6, 141980 Dubna}, \country{Russia}}}
\address[2]{\orgdiv{ExtreMe Matter Institute EMMI}, \orgname{GSI Helmholtzzentrum  f\"{u}r Schwerionenforschung}, \orgaddress{\state{ Planckstra\ss{}e 1, 64291  Darmstadt}, \country{Germany}}}
\address[3]{\orgdiv{Institute of Theoretical Physics}, \orgname{Wroclaw University}, \orgaddress{pl. M. Borna 9, 50-204 Wroclaw}, \country{Poland}}
\address[4]{\orgdiv{}\orgname{National Research Nuclear University (MEPhI)}, \orgaddress{\state{Kashirskoe shosse, 31, 115409 Moscow}, \country{Russia}}}
\address[5]{\orgdiv{Institut f\"{u}r} Kernphysik, \orgname{Technische  Universit\"{a}t Darmstadt}, \orgaddress{\state{ Schlossgartenstra\ss{}e 9, D-64289, Darmstadt}, \country{Germany}}}
\address[6]{\orgdiv{}\orgname{GSI Helmholtzzentrum  f\"{u}r Schwerionenforschung}, \orgaddress{\state{Planckstra\ss{}e 1, 64291  Darmstadt}, \country{Germany}}}

%%%%%%%%%%%%%%%%%%%%%%%%%%%%%%%%%%%%%%%%%

\corres{David E. Alvarez-Castillo, JINR, Joliot-Curie str. 6, 141980 Dubna, Russia . \email{alvarez@theor.jinr.ru}}

\presentaddress{Joliot-Curie str. 6, 141980 Dubna, Russia}

\begin{abstract}%
We present a multi-polytrope approach to describe high-mass twins fulfilling chiral effective 
field theory estimations of the neutron star equation state and test it against the appearance 
of mixed phases at the hadron-quark interface. 
In addition, we discuss astrophysical applications of this method and expected future 
measurements that shall further constrain neutron star matter and the understanding of the 
QCD phase diagram.  
\end{abstract}


\jnlcitation{\cname{%
\author{David Alvarez-Castillo,}
\author{David Blaschke}
and 
\author{Stefan Typel}} (\cyear{2017}), 
\ctitle{Mixed Phase Transition within the Multi-polytrope Approach to High Mass Twins} 
\cjournal{}\cvol{}.}

\maketitle

\section{Introduction}

Neutron stars (NS) are superdense compact objects with central densities
of the order of several times the nuclear saturation density $n_0=0.15$fm$^{-3}$, the mean 
density of atomic nuclei. 
Hence, exotic states of matter could exist in the interior of NSs. 
The elucidation of the interior composition of compact stars is of special importance when 
considering the QCD phase diagram. 
In case of a strong first-order phase transition from hadronic matter into any type of exotic 
matter, like quark deconfinement, the so-called \textit{High Mass Twin} (HMT) phenomenon
\citep{Alvarez-Castillo:2013cxa,Blaschke:2013ana,Benic:2014jia,Alvarez-Castillo:2015xfa,Kaltenborn:2017hus,Alvarez-Castillo:2015rwi} 
predicts particular characteristics of macroscopic observables in compact stars: disconnected 
sequences (families) in the mass-radius diagram featuring compact star branches with
overlapping ranges in the gravitational mass $M$ but with different ranges of radii so that
the radius difference at equal mass can vary from one half to a couple of kilometers, depending 
on the model description.
HMTs allow for a resolution of several problems in the description of dense nuclear matter 
and its relation to compact stars: the masquerade effect, the reconfinement EoS case and 
the hyperon puzzle~\citep{Blaschke:2015uva}. 
Moreover, HMTs lead to prediction of various astrophysical phenomena like energetic emissions 
that can potentially be detected. 
Furthermore, the recently detected gravitational wave emissions from the fusion of compact 
objects in binaries brings up the possibility of probing neutron star interiors. 
In this respect, NS-NS-merger calculations as performed, e.g., 
in~\citep{Bauswein:2014qla,Hanauske:2016gia} represent an important tool for  understanding 
these mergers.

\begin{figure}[!htb]
\centering
\SPIFIG{\includegraphics[width=0.5\textwidth, angle=0]{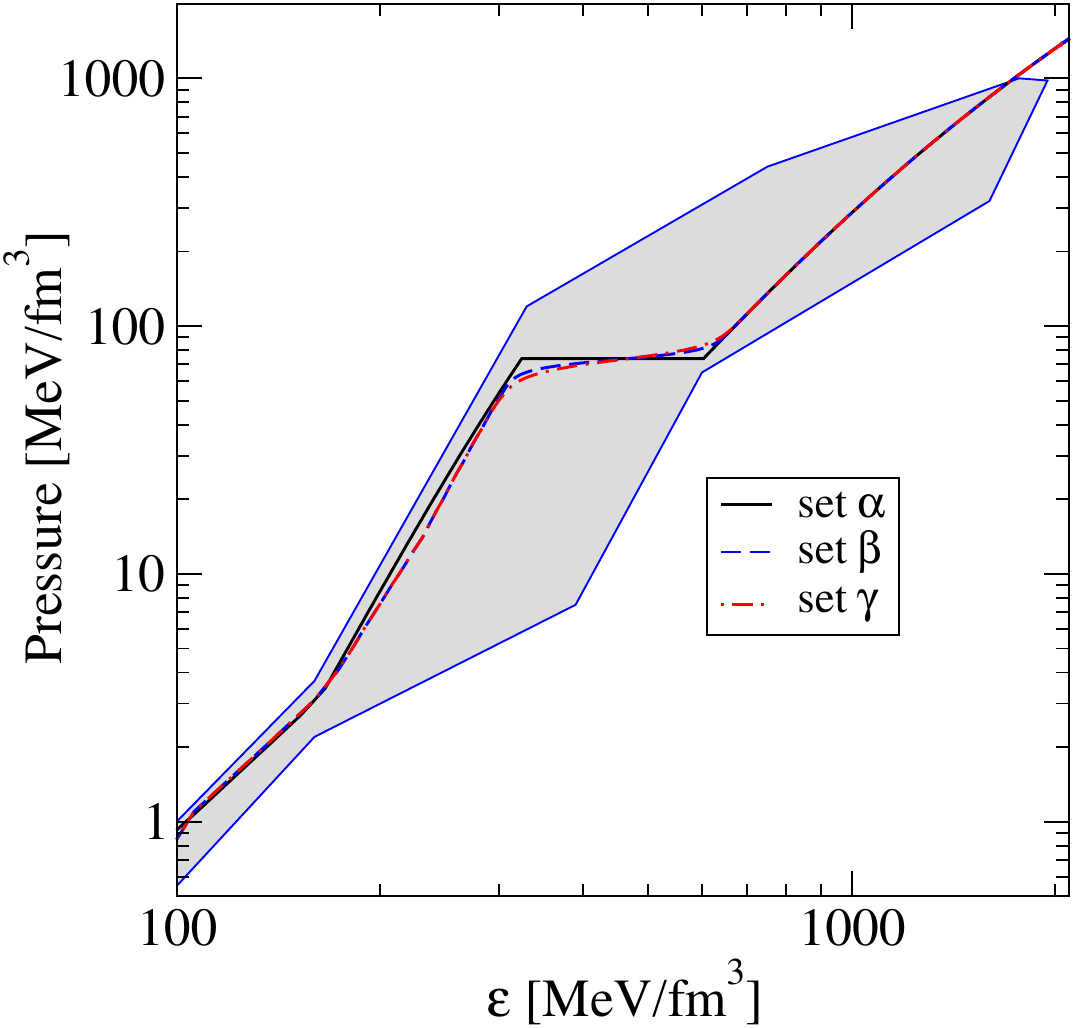}}
{\caption{Multipolytrope EoS sets within the Hebeler region~\citep{Hebeler:2013nza}.
\label{MP-Hebeler-pasta}}}
\end{figure}

The aim of this work is to show that within the \textit{multipolytrope approach} for the compact 
star equation of state (EoS)\textemdash see~\citep[and references therein]{Alvarez-Castillo:2017qki}\textemdash 
yields HMTs which are in wide limits robust against the appearance of pasta phases at the 
interface where a strong first order phase transition occurs. 
Without loss of generality, the multi-polytrope EoS consists of several pieces that fulfill 
$P_i(n)=\kappa_i n^{\Gamma_i}$, where $P$ is the pressure and $n$ the baryonic density. 
The coefficients $\kappa_i$ and $\Gamma_i$ serve to describe the state of matter in each
density region. 
As shown in~\citep{Alvarez-Castillo:2017qki} this approach is in accordance with the scheme 
of Hebeler et al.~\citep{Hebeler:2013nza} that constrains the high-density EoS with the 
compact star maximum mass constraint $M_{\rm max} \ge 2~M_\odot$ and at subnuclear densities
is in accordance with EoS constraints from chiral effective field theory, see 
figure~\ref{MP-Hebeler-pasta}.

The first necessary criterion for HMTs to appear is the so called Seidov 
constraint~\citep{1971SvA....15..347S}: 
\bea
\label{seidov}
\frac{\Delta\varepsilon}{\varepsilon_{\rm crit}} \ge \frac{1}{2} 
+ \frac{3}{2} \frac{P_{\rm crit} }{\varepsilon_{\rm crit}} 
\eea
that will lead to an instability in the M-R relation of compact stars. 
Within our multipolytrope description this can be easily achieved by 
defining one of the polytropes as a constant pressure region with 
$P=P_{\rm crit}$ with a jump in energy density $\Delta\varepsilon$ at the critical energy 
density $\varepsilon_{\rm crit}$ due to the strong first order phase transition. 
The second necessary condition is that the EoS at high densities is stiff enough to prevent
gravitational collapse and at the same time does not violate the causality constraint for the 
speed of sound $c_s<c$.

The corresponding star sequences are obtained by solving the Tolman-Volkoff-Oppenheimer (TOV) equations that describe a static, non-rotating, spherically symmetric star
\citep{Tolman:1939jz,Oppenheimer:1939ne}
\bea
\frac{dP( r)}{dr}&=& %\frac{-G }{r}
-\frac{G\left(\varepsilon( r)+P( r)\right)
\left(M( r)+ 4\pi r^3 P( r)\right)}{r\left(r- 2GM( r)\right)},\\
\frac{dM( r)}{dr}&=& 4\pi r^2 \varepsilon( r),
\eea
and by considering $P(r=R)=0$ and $ P_c= P(r=0)$ as boundary conditions for a compact star with mass $M$ and radius $R$, respectively. 

By increasing the chosen central pressure $P_c$ up to a value for which the maximum mass is reached, the complete compact star sequence is determined. Furthermore, the enclosed baryonic mass can be obtained by integrating
\bea
\frac{d N_B( r)}{dr}&=& 4\pi r^2 \left(1-\frac{2GM( r)}{r}\right)^{-1/2}n( r)~.
\eea
This quantity is of particular importance because it plays an important role in the dynamics 
of compact star evolution scenarios, like mass gain by accretion leading to a transition 
from a pure hadronic star into a hybrid star configuration \citep{Bejger:2016emu}.
In addition, a pure hadronic star whose central density is near the phase transition value 
could spin down into the hybrid twin configuration, a process expected to conserve the 
baryonic mass.

\section{High mass twins with mixed phase formation}
\begin{figure*}[!htb]
\SPIFIG{\includegraphics[width=0.95\textwidth, angle=0]{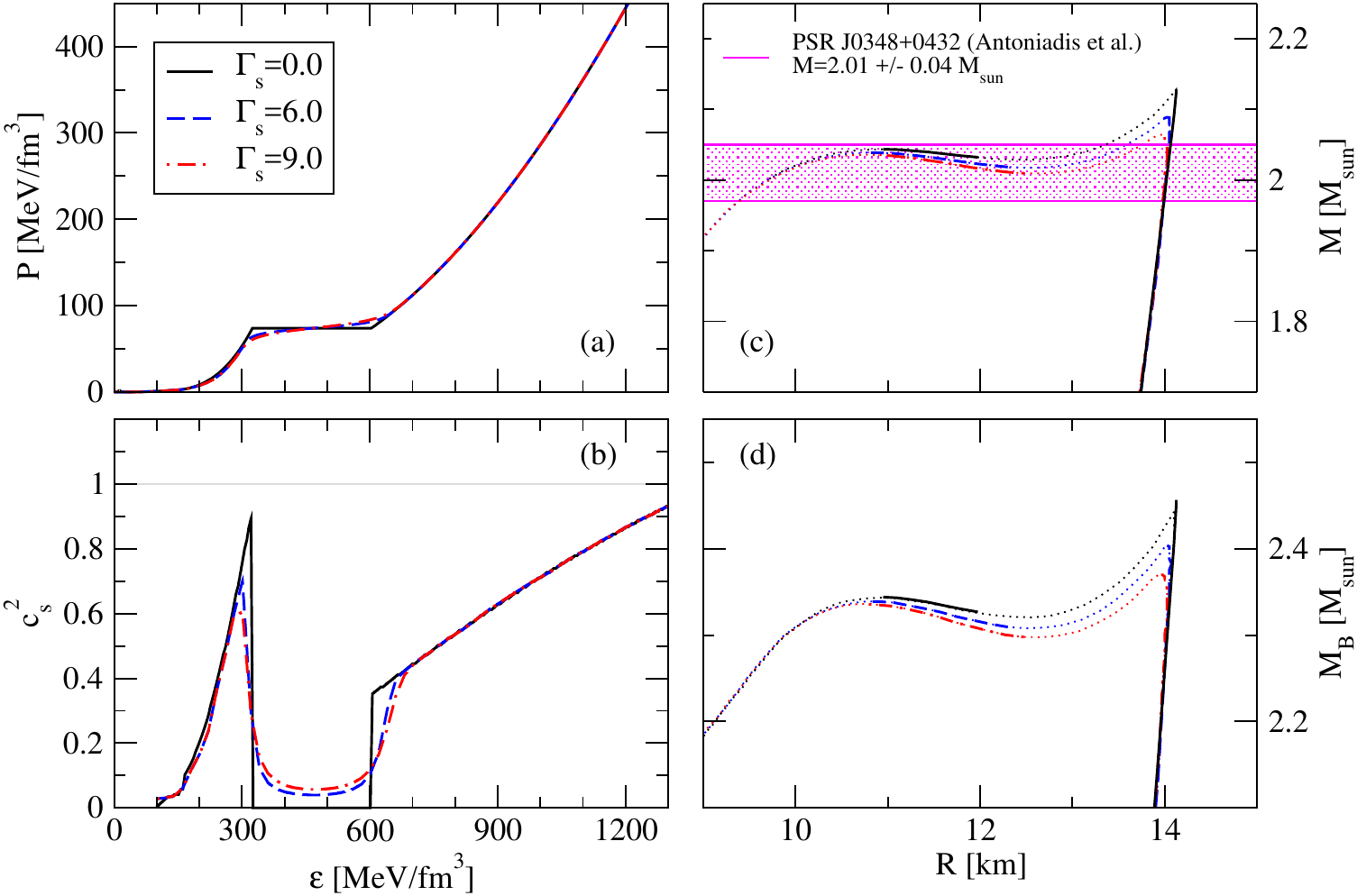}}
{\caption{EoS and sequences of compact stars. Mixed phase at the hadron-quark transition for model parameters of table~\ref{param-Pasta}.
\label{MP-Twins_pasta}}}
\end{figure*}

Pasta phases can appear at the hadron-quark boundary. By choosing an EoS for the HMT's (see~\citep{Alvarez-Castillo:2017qki} for details), the inclusion of such a mixed phase can be phenomenologically descibed by an interpolation. Here we adopt
a parametrization of the EoS of the form~\citep{Alvarez-Castillo:2014dva}:
\begin{equation}
 \varepsilon( p)=\varepsilon_h( p)f_<( p)+\varepsilon_q( p)f_>( p)~,
\end{equation}
\begin{equation}
f_{\lessgtr}(p) = \frac{1}{2} \left[1 + \textrm{tanh} \left(\mp \frac{p-p_c}{\Gamma_s}\right)\right]~,
\end{equation}
where $\varepsilon_h(p)$ and $\varepsilon_q(p)$ are the energy densities in the hadronic 
and quark matter phases, respectively. The resulting EoS are presented in 
figure~\ref{MP-Hebeler-pasta} where the parameter $\Gamma_s$ determines
the strength of the mix. 
The net effect is both smoothing the phase transition plateau with a non-zero slope resembling 
the $\Gamma_2\neq0$ multipolytrope case (see~\citep{Zdunik:2005kh,Read:2008iy}] for examples 
of such configurations).  
Panel (c) of Figure~\ref{MP-Twins_pasta} shows the high-mass part of the corresponding 
mass-radius sequences for the mixed phase parameters of table~\ref{param-Pasta}. 
One interesting effect of the mixed phase is the lowering of the speed of sound (panel (c)) 
at the critical density and resulting in a lowering of the maximum mass of the second branch. 
As a result, the baryonic mass difference for the twins is also reduced (see panel (d) 
in figure~\ref{MP-Twins_pasta}).

\begin{table}[htbp!]
\centering
\processtable{Parameter values for mixed phase sets with high density polytrope parameters $\Gamma_3=3.2$ and $\kappa_3=490.41$ MeV fm$^{3(\Gamma_3-1)}$.}{\tabcolsep=0pt%
\label{param-Pasta}
%\processtable{Double column table caption. Double column table caption. Double column table caption. Double column table caption. Double column table caption.\label{tab1}}
\begin{tabular}{l|c|ccc}
\hline \hline
& $\Gamma_s$ & $M_{\rm max}^{NS}$  & $M_{\rm max}^{HS}$ & $M_{\rm min}^{HS}$\\		
& [MeV fm$^{-3}$]  &[M$_{\odot}$] & [M$_{\odot}$] & [M$_{\odot}$] \\		
\hline
set $\alpha$& 0.0 &2.176 &~2.054 &~2.050 \\
set $\beta$& 6.0 &2.088 &~2.039 &~2.017 \\
set $\gamma$& 9.0 &2.064 &~2.035 &~2.009 \\
\hline \hline
\end{tabular}}
{\begin{tablenotes}%%[341pt]
%\footnotetext[]{Source: Example for table source text.}
\end{tablenotes}}
\end{table}

%\begin{figure}[t]
%\SPIFIG{\includegraphics[height=10pc,width=78mm]{blankfig}}{\caption{This is an example for appendix figure.\label{fig5}}}
%\end{figure}

\section{Conclusions}\label{sec5}

The multi-polytrope approach to the HMTs proves to be an effective tool for exploring the 
compact star EoS through diverse astrophysical phenomena. 
In this work we have found that the HMTs in this approach are robust against the appearance 
of pasta phases at the hadron-quark boundary.
Due to the flexibility in the parameter variation of the multi-polytrope approach it is 
feasible to perform a Bayesian Analysis for parameter estimates fulfilling observational 
constraints, as presented in various works~\citep{Ayriyan:2017nhp,Alvarez-Castillo:2016oln,Alvarez-Castillo:2015via,Ayriyan:2015kit,Blaschke:2014via,Alvarez-Castillo:2014nua,Raithel:2016bux,Raithel:2017ity}.
In addition, the recently deployed NICER detector is expected to measure the NS radius to 
an accuracy of $5\%$ for the first three NS candidates, therefore being potentially able to 
elucidate HMTs.
Moreover, gravitational wave signals from NS-NS mergers have been already studied 
in~\citep{Hanauske:2016gia} by using a multi-polytrope description with generic temperature 
addition but so far not including the HMTs case. 
The observation of such events by advanced LIGO is expected soon.
Further work in this direction is a natural step for testing the compact star
EoS and for exploring the QCD phase diagram. 
The resulting study will serve as a guide for future gravitational signal detection of this 
process in interferometers like advanced LIGO or VIRGO.
Last but not least, the SKA array in the southern hemisphere will gather pulsar data that 
shall constrain NS moments of inertia, masses and rotational values, all of them important 
quantities in the study of the NS EoS.

\begin{bm}[Acknowledgments]
 D.E.A-C. and S.T. received support from the Heisenberg-Landau programme.
D.B. was supported in part by the MEPhI Academic Excellence Project under grant 
No. 02.a03.21.0005 and by the Polish National Science Centre (NCN) under grant number 
UMO - 2014/13/B/ST9/02621. D.E.A-C is grateful for support
from the programme for exchange between JINR Dubna and Polish Institutes (Bogoliubov-Infeld programme). 
This research was supported in part by the ExtreMe Matter Institute EMMI at the GSI 
Helmholtzzentrum fuer Schwerionenphysik, Darmstadt, Germany.

\subsection*{Author contributions}

All authors contributed equally to the elaboration of this article.
\subsection*{Financial disclosure}

None reported.

\subsection*{Conflict of interest}

The authors declare no potential conflict of interests.
\end{bm}

\section*{References}

\nocite{*}
\bibliography{wileyarticle}%

%\begin{biography}{{\includegraphics[width=5pc,height=5pc]{blankfig.eps}}}{\textbf{Author Name.} This is author biography text. This is author biography text. This is author biography text. This is author biography text. This is author biography text. This is author biography text. This is author biography text. This is author biography text. This is author biography text. This is author biography text.}
%\end{biography}
\end{document}